\documentstyle[prl,aps]{revtex}
\input{epsf}
\draft
\begin{document}
\twocolumn[\hsize\textwidth\columnwidth\hsize\csname@twocolumnfalse\endcsname
\title{Dynamics of Wetting Fronts in Porous Media}
\author{Igor~Mitkov}
\address{
Center for Nonlinear Studies and
Computational Science Methods Group\\
Los Alamos  National Laboratory,
Los Alamos, NM 87545}
\author{Daniel~M.~Tartakovsky and C.~Larrabee~Winter}
\address{
Geoanalysis Group, Earth and Environmental Science Division\\
Los Alamos  National Laboratory,
Los Alamos, NM 87545
}
\date{\today}
\maketitle

\begin{abstract}
We propose a new phenomenological approach for describing
the dynamics of wetting front propagation in porous media.
Unlike traditional models, the proposed approach is based 
on dynamic nature of the relation between capillary pressure
and medium saturation.
We choose a modified phase-field model of solidification
as a particular case of such dynamic relation.
We show that in the traveling wave regime the results obtained
from our approach reproduce those derived from the standard model of
flow in porous media. In more general case, the proposed
approach reveals the dependence of front dynamics upon the flow regime.
\end{abstract}


\narrowtext
\vskip1pc]

The dynamics of fluids in porous media has been a subject
of numerous theoretical and experimental
studies because of its importance for engineering and environmental
applications~\cite{sahimi,barenblatt,scheidegger}.
Among the most challenging problems in this area is
modeling fluid flow through a partially saturated medium,
in particular the propagation of wetting (or drying) fronts.
This has been addressed on both ``microscopic'' (or pore-scale) and
``macroscopic'' (on scales larger than pore size) levels.
The fluid dynamics in the pore networks has been studied
by numerous authors both theoretically~\cite{theory}
and experimentally~\cite{lenormand,exper}.
These studies on the microscopic
scale provide valuable insight into the underlying mechanisms
of fluid transport in porous media.
A microscopic description
requires the detailed information of the pore structure
and pore-size distribution.
When this information is not available, as often happens
for large domains, one has to rely upon a macroscopic description.
This description is useful for determining such integral
characteristics, as the
width and propagation velocity of moving wetting fronts,
the influence of its curvature on the dynamics, etc.

The subject of the present work is the dynamics of wetting
fronts in porous media when a liquid phase (water) displaces
air. A straightforward description of this
process consists of treating wetting fronts as sharp interfaces
which separates completely wet and dry regions~\cite{lenormand,kadanoff}.
However, in many realistic cases the structure of the transitional
zone (wherein water saturation varies gradually) cannot be
neglected. One can either describe the dynamics of the liquid and
air phases separately, or noting that air pressure is close
to atmospheric, consider only the dynamics of the liquid phase.
These two approaches generally produce similar results (Ref.~\cite{marsily},
p. 213). The latter approach seems to be more attractive due to its
simplicity.

Within the framework of the chosen approach water flow is described by
Darcy's law which, similarly to Ohm's law for electric current,
stipulates the proportionality between flux and gradient
of a potential (see, {\it e.g.},~\cite{kadanoff}).
For flow through porous media a water pressure,
averaged over many pore sizes, plays the role of the potential.
Richards~\cite{richards} has empirically
generalized Darcy's law onto flow in partially saturated porous media
(PSPM) by letting the proportionality coefficient depend on saturation
$\theta$ of the medium ($0 < \theta < 1$). Coupled with the mass
conservation law, the generalized Darcy's law constitutes
the existing macroscopic description of the fluid dynamics in
PSPM. To complete this description, one needs to specify
a relation between $\theta$ and capillary pressure
(normalized by the product of fluid density
$\rho$ and gravitational acceleration $g$), $\psi$.
Traditionally this relation is assumed
to be algebraic~\cite{sahimi,bear}. However, this contradicts the numerous
experimental evidences revealing the dependence of
$\theta - \psi$ relation
upon the conditions of the experiment. In particular, this
relation exhibits {\em hysteresis} for wetting and drying of
a medium~\cite{sahimi,bear,miller}.

In the present Letter we propose a novel phenomenological approach
to describe propagation of wetting fronts in porous media,
which is based on a {\em dynamic} $\theta-\psi$
relation. Our description consists of two dynamic equations for $\theta$
and $\psi$ coupled by nonlinear sources.
This implies that the traditional algebraic $\theta-\psi$
correspondence
is replaced by a nonlocal (integro-differential) relationship.
The proposed model is a new application of the well-known
``phase-field'' approach used to describe solidification,
electro-deposition, and other physical
problems~\cite{karma,karma1,langer,collins,kobayashi,wang}.
We show that, under certain conditions, dynamics of the wetting
front in our approach reproduce that in the traditional approach.
The same is true for the pressure profiles associated with the wetting
fronts obtained from both models.
We demonstrate that, under different conditions, this equivalence
breaks down, with our model revealing a dynamic nature
of $\theta-\psi$ relation.

The generalized Darcy's law for flux ${\bf q}$ through PSPM has a form
${\bf q} = -K(\theta) \nabla (\psi - x_3)\,,$ where $K$ is the
saturation-dependent conductivity of the medium and $x_3$ is
the vertical coordinate (positive downward) that stands for
the gravitational component of pressure (normalized by $\rho g$).
Here $\psi$ is measured relative to the ambient atmospheric
pressure, so that $\psi < 0$ for partially saturated and
$\psi \geq 0$ for fully saturated medium.
Combined with the mass conservation law, $\partial\theta/\partial t =
-\nabla\cdot {\bf q}\,,$ this leads to the equation introduced
by Richards~\cite{richards}
\begin{equation}
\frac{\partial\theta}{\partial t} = \nabla\cdot [\,K(\theta)
\nabla (\psi - x_3)\,]\;.
\label{eq1}
\end{equation}
Over the years a number of algebraic $\theta-\psi$ and $K(\theta)$
relations have been proposed empirically (see, {\it} e.g.
\cite{gardner,vangen,mualem}).
Richards equation (RE)~(\ref{eq1}) combined with such a relation
constitutes the traditional approach to describe flow in PSPM.
Since algebraic relations are not supported by experimental
data, in what follows we propose a new approach with
a dynamic $\theta-\psi$ relation.

To describe flow in PSPM, we modify the phase-field model (PFM)
from~\cite{karma}
\begin{eqnarray}
\frac{\partial \psi}{\partial t} \;&=&\; D\nabla^2 \psi
\;-\; \frac{1}{S}\,\frac{\partial\theta} {\partial t}
\;-\; D\frac{\partial\theta}{\partial x_3}\;,
\label{eq3}\\
\tau \frac{\partial \theta}{\partial t} \;&=&\; W^2 \nabla^2 \theta
\label{eq4}\\
\;&+&\; \left[\; 2\theta - 1 - \lambda (\psi-\psi_f)\theta(1-\theta)\;\right]
\;\theta(1-\theta)\;,
\nonumber
\end{eqnarray}
where $D = K_s/S$ is the diffusion coefficient, $K_s$ is the conductivity
of a fully saturated medium, $S$ is the specific storage (measure
of compressibility of the fluid and medium),
$\tau$ is a characteristic time-scale of the saturation dynamics,
and $W$ is the width of a moving wetting front.
The model parameter $\lambda$ will be determined
as a function of macroscopic parameters $K_s\,,\, W\,$ and $\tau\,$.
The constant $\psi_f$ is the normalized capillary pressure along
the moving front in the sharp-interface limit.
We will demonstrate below that $\psi_f$ coincides with the
pressure in the dry medium far ahead of the wetting front.
Since the width of the capillary zone,
associated with the localized wetting fronts,
is much smaller than the typical scale of the pressure variation,
the following holds $W^2/\tau \ll D$.
We have added the last term in (\ref{eq3}) to PFM~\cite{karma},
to account for the gravitational force.

Our model is phenomenological in the sense that presently
we do not provide a rigorous physical motivation for the nonlinear
source term on the right-hand-side of (\ref{eq4}).
Nevertheless, the model captures the main
features of the wetting fronts propagation in porous media.
In particular, numerous experiments~\cite{front,haverkamp} have shown that,
under certain conditions, the wetting fronts remain localized
and propagate in a self-similar manner.
The medium is fully saturated ($\theta = 1$) behind the front region
and completely dry ($\theta = 0$) ahead of the front.
The cubic polynomial in (\ref{eq4}) provides for such a structure
of the wetting front (similar to the PFM for solidification).
The fact that the liquid moves in the direction opposite
to the pressure gradient is accounted for by the term proportional
to $(\psi-\psi_f)$ in~(\ref{eq4}), since its presence
makes the depths of the two minima of the corresponding potential energy
different.

Eqs. (\ref{eq3})--(\ref{eq4})
have to be supplemented by a constraint to ensure conservation
of mass. The specific storage is related to the compressibility
of fluid and porous medium as
$S = \rho g \omega (\beta_f - \beta_s + \beta_p)\,$ (Ref.~\cite{marsily},
p.~108).
Here $\omega$ is the medium porosity (fraction of pore volume in the
total volume $v$ of the medium) and
$\beta_f, \beta_m\,,$ and $\beta_p$ are the compressibility coefficients
of fluid, solid grains, and pores, respectively.
The total mass of the fluid is given by $M = \int_{v}\,\omega
\,\rho\,\theta\,d{\bf x}\,$.
For incompressible fluids and media
($S = 0$), since $D = K_s/S$, Eq. (\ref{eq3})
reads $\partial\theta/\partial t = K_s(\nabla^2\psi - \partial\theta/
\partial x_3)\,,$ which conserves mass.
For compressible fluids and media
$d M/dt = Q$, where $Q$ is the total mass flux through the medium,
provides the global constraint on the system (\ref{eq3})--(\ref{eq4}).

Since $S$ is the {\em specific} storage of a medium,
the effect of compressibility of fluids and media on flow
dynamics is characterized by the dimensionless parameter
$S L$, where $L$ is a typical domain size.
For many practical applications, such as flow of water
through low porosity rocks, $S \sim 10^{-5} - 10^{-7} m^{-1}$ and
$L \sim 10^{-1} - 10^{3} m$. Thus $S L \ll 1$, and the fluid
and medium are virtually incompressible.

We consider propagation of one-dimensional (1D) wetting fronts
in a slightly compressible medium with $S L \ll 1$.
The dynamics of these fronts is described by 1D
Eqs. (\ref{eq3})--(\ref{eq4}), and the mass conservation
constraint is satisfied automatically.
Two different situations exist: horizontal (gravity-free)
and vertical front propagation.
To maintain propagation of a self-similar front, we
choose the constant flux condition $\partial(\psi - x_3)/
\partial x_i = -q/K_s$ at $x_i = 0$, and no-flux condition
for capillary force $\partial \psi /\partial x_i = 0\,$
at $x_i = L$.
Here $i=1$ or $i=3$ for horizontal or vertical flow,
respectively, and $q$ is an external velocity flux.
At the boundary behind the front, $x_i = 0$, the medium
is fully saturated ($\theta = 1$), and at the boundary ahead
of the front, $x_i = L$, the medium is dry ($\theta = 0$).

Let us apply a traveling wave ansatz $z = x_i - V t$, for
a front moving with velocity $V$, to
1D Eqs. (\ref{eq3})--(\ref{eq4}).
Relations $W \sim V\tau$ and $W^2/\tau \ll D$ give rise to a small
parameter $Pe \ll 1$, where $Pe = VW/D$ is the P\'{e}clet number.
The perturbation analysis around $Pe = 0$, similar to that performed
in~\cite{karma}, gives a front moving with velocity $V = q$,
and saturation and capillary pressure profiles
\begin{eqnarray}
\theta(z) \;&=&\; \frac{1}{2}\,\left[1 - \tanh
\left(\frac{z}{2\sqrt{2}W}\right) \,\right] \;+\; O(Pe)\;,
\label{eq5}\\
\psi(z) \;&=&\; \psi_f\;+\; J\,\left\{\,W\sqrt{2}\ln{2}\;-\;
\frac{z}{2}\right.
\label{eq6}\\
\;&+&\; \left. W \sqrt{2} \ln
\left[\cosh\left(\frac{z}{2\sqrt{2}W}\right)\,\right]\,\right\}
\;+\; O(Pe^2)\;.
\nonumber
\end{eqnarray}
Here $J$ is an absolute value of the capillary pressure gradient
at the boundary.
It is given by $J = q/K_s$ for a horizontally propagating front,
and by $J = (q/K_s-1)$ for a vertically propagating front.
A vertical wetting front does not exist when $q \leq K_s$, {\it i.e.}
when the external flux is not large enough to compensate the effect
of the gravity force, and thus develop a saturated zone.

The saturation profile (\ref{eq5}) is evaluated in the
zeroth order in $Pe$, since for the localized wetting fronts
saturation varies in the narrow region $W \ll \sqrt{D\tau}$.
Hence the higher order corrections to $\theta$ influence the
nonlocal pressure profile (\ref{eq6}) only in the second order.
The solvability condition for the equation for $\theta$
in the first order (similar to~\cite{karma}) yields
$\tau/\lambda \approx -0.313 W^2 J/q\,$.

\begin{figure}[h]
\hspace{3.0cm}
\rightline{ \epsfxsize = 10.0cm \epsffile{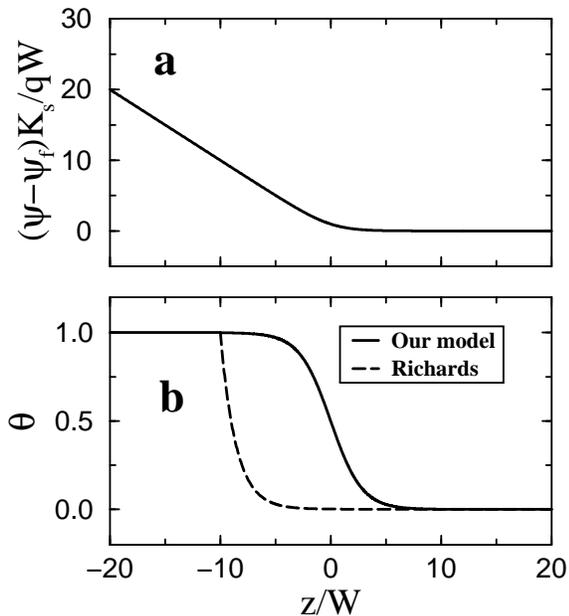}}
\caption{
Normalized capillary pressure $(\psi-\psi_f)K_s/q W$ (a)
and saturation (b) profiles in traveling wave coordinate.
In (b) normalized capillary pressure in a dry medium is taken
$\psi_f K_s /q W = - 10\,$.
\label{fig1}}
\end{figure}

We now compare our approach with the traditional
(Richards) approach. One of the most widely-used examples
of algebraic functions $\theta(\psi)$ and $K(\theta)$, complementary
to RE (\ref{eq1}), has been proposed by Gardner~\cite{bear,gardner}
\begin{equation}
\theta(\psi) = e^{\alpha\psi}\;,\;\;\;\;\;\;\;
K(\theta) = K_s \theta\;,
\label{eq2}
\end{equation}
where $\alpha$ is the pore-size distribution parameter.
The first equation in (\ref{eq2}) is valid for $\psi < 0$
(when the medium is partially saturated), while for
$\psi \geq 0$ (for a fully saturated medium) $\theta \equiv 1\,$.
The same boundary conditions as before are used. 
Substituting (\ref{eq2}) into 1D Eq. (\ref{eq1})
and applying the traveling wave ansatz gives, after
a series of transformations, the solution for the pressure
profile
\begin{eqnarray}
\psi(z) = \frac{\ln{2}}{\alpha} + \psi_L - \frac{J z}{2}
+ \frac{1}{\alpha} \ln\left[\cosh\left(\frac{\alpha J}{2}\,z\right)
\right] \;,
\label{eq7}
\end{eqnarray}
where $\psi_L$ is a large negative number corresponding to
the pressure in the dry medium far ahead of the front.

Comparing (\ref{eq7}) with (\ref{eq6}) we find that the pressure
profiles obtained from both models coincide, provided
the relations between parameters $\sqrt{2}WJ = \alpha^{-1}$
and $\psi_f = \psi_L$ hold. Figure~\ref{fig1}(a) shows
the resulting pressure profile from both models.
Substituting (\ref{eq7}) into the first relation in (\ref{eq2})
we obtain the saturation profile for Richards model. The result
is compared with the saturation profile (\ref{eq5}) in the
Figure~\ref{fig1}(b). Although the saturation profiles
differ within the capillary zone, this difference occurs
on the very small scale $W\,$. An unphysical sharp kink
that appears in the Richards model for $\theta(z)$
is absent in our model.
Comparing Figure~\ref{fig1}(b) with the experimental data
presented on Figures~2-5 of Ref.~\cite{haverkamp} shows that the saturation
profile obtained by our model agrees with the experiments.

\begin{figure}[h]
\hspace{-1.0cm}
\rightline{ \epsfxsize = 8.0cm \epsffile{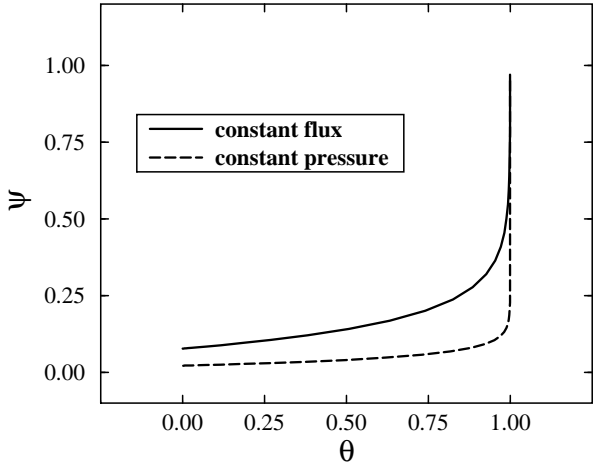}}
\caption{Dynamic $\theta-\psi$ relation for both traveling wave
and non-traveling wave regime.
Parameters of the simulations are $D = 1\,,W = 0.1\,,
S = 0.005\,,\tau = 1\,$. For the traveling wave
regime $q = K_s\,$. Value of $\lambda$
is calculated according to the selection result
$\tau/\lambda \approx -0.313 W^2 J/q\,$. System size
$L = 10\,$ and number of grid-points is 201. Time step
$dt = 0.001\,$.
\label{fig2}}
\end{figure}

Though in the traveling wave regime our model reproduces
the results obtained from the Richards model, it is also
capable of capturing the dynamic $\theta-\psi$ relation
in different regimes, without adjusting the parameters.
This is not the case for the Richards model, where one needs
to adjust the parameter $\alpha$ in (\ref{eq2}) to different
experimental regimes.
To demonstrate the dynamic nature of the $\theta-\psi$ relation
in our model, we have performed the numerical simulations of
1D Eqs. (\ref{eq3})--(\ref{eq4}) under several boundary
conditions. We solved system (\ref{eq3})--(\ref{eq4})
using finite differences.
Note that though in our simulations $S L \ll 1$, we keep
the term $S \partial\psi/\partial t$ in (\ref{eq3}). Despite
being small, this term provides relaxational feature
to numerical solution of (\ref{eq3}), which stabilizes
the numerical algorithm.
Figure~\ref{fig2} shows $\theta(\psi)$ obtained
from these simulations for horizontal flow. The solid line in
the figure corresponds to
the traveling wave regime, resulting from the previously
described boundary conditions. The dashed line
represents $\theta(\psi)$ at time $t=1000$ corresponding
to the wetting with the constant pressure at $x_1=0$ and
the same no-flux condition at $x_1 = L\,$.

Note that in the limit of the narrow capillary zone,
$W\rightarrow 0\,$, our model describes the dynamics of sharp interface
which separates completely wet ($\theta=1$) and completely dry
($\theta=0$) regions.
This is in analogy with the PFM of solidification
that reduces to the free-boundary problem~\cite{karma}.
It follows from (\ref{eq6}) that in this limit $\psi(0) = \psi_f$ and
$\psi_z(0) = -J\,$. Moreover, (\ref{eq3})
reduces to a diffusion equation which is commonly used to 
describe flow in saturated media.

In conclusion, we have proposed a novel approach to describe
dynamics of wetting fronts in porous media.
Unlike the traditional approach, our phenomenological approach reflects
the dynamic nature of the relation between capillary pressure
and saturation of a medium.
We have found that this relation varies with the flow regimes,
which is supported by experimental data.
We have demonstrated that, in the traveling wave regime, the proposed model
reproduces the results obtained from the standard approach.
We plan to extend our novel approach to two- and three-dimensional
media incorporating in the description the front curvature and
anisotropy of the medium.
We expect to develop experimental support for the proposed approach
by measuring quantitative features of the wetting fronts in
porous media, such as the dependence of the front width on
external flux.

We are grateful to R. Camassa, G. Forest, M. Hyman, S.-Y. Chen, and J. Glazier
for fruitful and clarifying discussions.

\end{document}